# A machine learning approach to drug repositioning based on drug expression profiles: Applications to schizophrenia and depression/anxiety disorders

Kai Zhao[1] and Hon-Cheong So[*,1,2]

[1]School of Biomedical Sciences, The Chinese University of Hong Kong, Shatin, Hong Kong

[2]KIZ-CUHK Joint Laboratory of Bioresources and Molecular Research of Common Diseases, Kunming Zoology Institute of Zoology and The Chinese University of Hong Kong

Corresponding author: Hon-Cheong So. Email: hcso@cuhk.edu.hk

**Abstract**

Development of new medications is a very lengthy and costly process. Finding novel indications for existing drugs, or drug repositioning, can serve as a useful strategy to shorten the development cycle. In this study, we present an approach to drug discovery or repositioning by predicting indication for a particular disease based on expression profiles of drugs, with a focus on applications in psychiatry. Drugs that are not originally indicated for the disease but with high predicted probabilities serve as good candidates for repurposing. This framework is widely applicable to any chemicals or drugs with expression profiles measured, even if the drug targets are unknown. It is also highly flexible as virtually any supervised learning algorithms can be used.

   We applied this approach to identify repositioning opportunities for schizophrenia as well as depression and anxiety disorders. We applied various state-of-the-art machine learning (ML) approaches for prediction, including deep neural networks, support vector machines (SVM), elastic net, random forest and gradient boosted machines. The performance of the five approaches did not differ substantially, with SVM slightly outperformed the others. However, methods with lower predictive accuracy can still reveal literature-supported candidates that are of different mechanisms of actions. As a further validation, we showed that the repositioning hits are enriched for psychiatric medications considered in clinical trials. Notably, many top repositioning hits are supported by previous preclinical or clinical studies. Finally, we propose that ML approaches may provide a new avenue to explore drug mechanisms via examining the variable importance of gene features.

## INTRODUCTION

Development of new medications is a very lengthy and costly process. While investment in research and development has been increasing, there is a lack of proportional rise in the number of drugs approved in the past two decades, especially for drugs with novel mechanisms of actions [1]. There is an urgent need for innovative approaches to improve the productivity of drug development. This is particularly true for some areas like psychiatry, for which there has been lack of therapeutic advances for some time [2,3].

Finding new indications for existing drugs, an approach known as drug repositioning or repurposing, can serve as a useful strategy to shorten the development cycle [4]. Repurposed drugs can be brought to the market in a much shorter time-frame and at lower costs. With the exponential growth of "omics" and other biomedical data in recent years, computational drug repositioning provides a fast, cost-effective and systematic way to identify promising repositioning opportunities [4].

In this study we describe a general drug repositioning approach by predicting drug indications based on their expression profiles, with a focus on applications in psychiatry. We treat drug repositioning as a supervised learning problem and apply different state-of-the-art machine learning methods for prediction. Drugs that are not originally indicated for the disease but have high predicted probabilities serve as good candidates for repositioning. There are several advantages of this approach. Firstly, the presented approach is a general and broad framework that leverages machine learning (ML) methodologies, a field with very rapid advances in the last decade. This provides great flexibility and opportunities for further improvement in the future as virtually any supervised learning methods can be applied. Newly developed prediction algorithms can also be readily incorporated to improve the detection of useful drug candidates. In addition, the method described here is widely applicable to any chemical or drugs with expression profiles recorded, even if the drug targets or mechanisms of actions are unknown. For example, herbal medicine products may contain a mixture of ingredients with uncertain drug targets; even for many known medications [*e.g.* lithium [5]], their mechanisms of actions and exact targets are not completely known. If transcriptomic profiling has been performed, they can still be analyzed for therapeutic or repositioning potential under the current approach.

There has been increasing interest in computational drug repositioning recently, in view of the rising cost of new drug development. Hodos et al.[6] provided a comprehensive and updated review on this topic. Similarity-based methods (*e.g.* ref.[7-12]) represent one common approach, but as noted by Hodos et al., the dependence on data in the "nearby pharmacological space" might limit the ability to find medications with novel mechanisms of actions. Another related methodology is the network-based approach [13], which typically requires data on the relationship between drugs, genes and diseases as well as connections within each category (e.g. drug-drug similarities). It can integrate different sources of information but may still be constrained by the focus on a nearby pharmacological space and the choice of tuning parameters in network construction or inference is often ad hoc[6]. The present work is different in that we offer a broad framework for repositioning and we do not focus on one but many different kinds of learning methods. There is comparatively less reliance on known drug mechanisms or the "nearby pharmacological space" as we let the different algorithms "learn" the relationship between drugs, genes and disease *in their own ways*. We note that kernel-based ML methods such as support vector machine (SVM) are also based on some sort of similarity measures, but here we considered a variety of other approaches and SVM is one of the methods which falls under the broader framework of ML for repositioning. For high-throughput omics data, often only a subset of genes or input features are relevant and a machine learning approach is able to "learn" which features to consider for repositioning. As we shall discuss later, ML approaches also provide a new avenue to explore the mechanisms of different drug classes, by studying the variable importance of input genes.

We are particularly interested in drug repositioning for psychiatric disorders in view of the lack of novel treatments in the area. It is worth noting that while psychiatric disorders are leading causes of disability worldwide[14], there have been limited advances in the development of new pharmacological agents in the last two decades or so[3]. Development of new therapies is also limited by the difficulty of animal models to fully mimic human psychiatric conditions[15]. Investment by drug companies has in general been declining[3], and new approaches for drug discoveries are very much needed in this field. We will explore repositioning opportunities for schizophrenia along with depression and anxiety disorders. Here depression and anxiety disorders are analyzed together as they are highly clinically comorbid[16,17], show significant genetic correlations[18], and share similar pharmacological treatments[19].

Contributions of this study are summarized below. Firstly, we presented a general approach to drug repositioning based on ML methods, leveraging drug expression profiles as predictors. While previous work [20] has also proposed the use of ML on drug transcriptome profiles for classifying drugs into groups (e.g. anti-cancer drugs, cardiovascular drugs, drugs acting on the central nervous system etc.), we focused on drug repositioning for *particular diseases* instead of predicting the big therapeutic groups. Secondly, we have performed a comparison of the predictive performances of five state-of-the-art and perhaps most commonly employed ML algorithms, including deep neural networks, support vector machines, elastic net, random forest and gradient boosted trees. Thirdly, we identified new repositioning opportunities for schizophrenia and depression/anxiety disorders, and validated the relevance of the repositioned drugs by showing their enrichment among drugs considered for clinical trials, as well as support by previous literature. We also found that methods with slightly lower predictive accuracy still reveal literature-supported repositioning candidates that are of different mechanisms of actions from the drugs listed by better-performing algorithms. Finally, we explored which genes and pathways contributed the most to our predictions, hence shedding light on the molecular mechanisms underlying the actions of antipsychotics and antidepressants.

**METHODS**

We present a general drug repositioning approach adopting a supervised learning approach. We construct prediction models in which the outcome is defined as whether the drug is a known treatment for the disease, and the predictors are expression profiles of each drug. Drugs that are *not* originally known to treat the disease but have high predicted probabilities are regarded as good candidates for repositioning.

**Drug expression data**
The expression data is downloaded from the Connectivity Map (CMap), which captures transcriptomic changes when thee cell lines (HL60, PC3, MCF7) were treated with a drug or chemical[21]. We downloaded raw expression data from Cmap, and performed normalization with the MAS5 algorithm[22]. Expression levels of genes represented on more than one probe sets were averaged. We employed the limma package[23] to perform analyses on differential expression between treated cell lines and controls. Analyses were performed on each combination of drug and cell line, with a total of 3478 instances. Expression measurements were available for 12436 genes. Statistical analyses were performed in R3.2.1 with the help of the R package "longevityTools".

**Defining drug indications**

Drug indications were extracted from two known resources, namely the Anatomical Therapeutic Chemical (ATC) classification system and the MEDication Indication Resource high precision subset (MEDI-HPS)[24]. We focus on schizophrenia as well as depression and anxiety disorders in this study. From the ATC classification system, two groups of drugs were extracted, including antipsychotics and antidepressants. On the other hand, the MEDI-HPS dataset integrates four public medication resources, including RxNorm, Side Effect Resource 2 (SIDER2)[25], Wikipedia and MedlinePlus. We used the MEDI high-precision subset (MEDI-HPS) which only include drug indications found in RxNorm or in at least 2 out of 3 other sources[24]. This subset achieves a precision of up to 92% according to Wei et al.[24].

**Machine learning methods**

We employed different state-of-the-art machine learning approaches including deep neural networks (DNN)[26], support vector machine (SVM)[27], random forest (RF)[28], gradient boosted machine with trees (GBM)[29] and logistic regression with elastic net regularization (EN)[30] to predict indications. Our data is imbalanced as only a minority of the drugs are indicated for schizophrenia or depression/anxiety disorders. We performed both unweighted and weighted analyses in this study; in the weighted analysis, class weights are adjusted such that the minority group (drugs indicated for the disorder) will receive higher weight to achieve a balance between positive and negative instances.

In our unweighted model, DNN was implemented in the python package keras. Hyperparameters were chosen by the "fmin" optimization algorithm from "hyperopt", which employs a sequential model-based optimization approach[31]. The tree-structured Parzen estimator (TPE) was used. The more sophisticated hyper-parameter search strategies provided by sequential model-based methods may produce better results than simpler approaches (e.g. grid search) when the number of hyper-parameters is large, such as in deep learning settings [31]. Fifty evaluations were run for each search of optimal hyper-parameters. Dropout and mixed L1/L2 penalties were employed to reduce over-fitting. The neural networks consisted of two or three layers, with number of nodes selected uniformly from the range [64, 1024]. Dropout percentage was selected uniformly from [0.25, 0.75], and L1/L2 penalty uniformly from [1E-5, 1E-3]. Optimizer was chosen from "adadelta"[32], "adam"[33] and "rmsprop"[34], and the activation function chosen from "relu", "softplus" or "tanh". One hundred epochs were run for each model and we extracted the model weights corresponding to the best epoch.

We also attempted 'hyperopt' for the weighted analysis, however the predictive performance was unexpectedly poor due to unclear reasons yet to be revealed. We therefore turned to hyperparameter selection with grid search, with some adjustments in the parameter ranges. A two-layer neural network was constructed with dropout and mixed L1/L2 to avoid over-fitting. The number of neurons in the first hidden layer was selected from {1000, 1500, 2000}, the number in the second layer from {500, 1000, 1500}, dropout rate from {0.4, 0.5, 0.6, 0.7, 0.8}, L1/L2 penalty from intervals [-13, -3] and [-9, -8] in log space and the number of epochs from [10, 20, 30, 50]. In order to speed up hyperparameter selection, we first chose the number of epochs, the best optimizer and activation function (following the same parameter range as described above) with other parameters fixed, and then used the best parameters chosen in the first step to find the optimal complexity of our neural networks by selecting the number of neurons in each layer, dropout and mixed L1/L2 penalties.

SVM, RF and GBM models were implemented in "scikit-learn" (sklearn) in python. Hyper-parameter selection was performed by the built-in function GridSearchCV in sklearn. For SVM, we chose radial basis function as the kernel. The two hyper-parameters C and gamma were selected from [-5, 2] and [-6, 2] in log10-space respectively.

For RF, the number of bagged trees was set to 1000, the maximum number of features used for splitting was selected from {800, 1000, 1500, 2000, 3000, 5000} and min_samples_leaf (the minimum number of samples required at a leaf node) was selected from {1, 3, 5, 10, 30, 50, 80}. For GBM, the number of boosting iterations was selected from a sequence of 100 to 1001 with step size 50, learning rate from {0.005, 0.01, 0.015, 0.02, 0.03, 0.05}, subsampling proportion from {0.8, 1}, maximum depth of each estimator from {2, 3, 5, 10} and maximum number of features from {10, 30, 50, 100, 500, 1000}. Finally, the EN model was implemented by the R package "glmnet"[35]. The elastic-net penalty parameter $\alpha$ was chosen from seq(0, 1, by=0.1), with other settings following the default.

**Nested Cross-validation**

We adopted nested three-fold cross validation (CV) to choose hyper-parameters and evaluate model performances. It has been observed that optimistic bias will result if one uses simple CV to compute an error estimate for a prediction algorithm that itself is tuned using CV[36]. Nested CV avoids this problem and is able to give an almost unbiased estimate of prediction accuracy[36]. The inner loop CVs were used to choose the parameters that optimized predictive performance. In each outer loop CV we made predictions on the corresponding test set using the best model trained from the inner CV loops. To achieve maximum consistency in our comparisons, we compared different methods on the same test set in each loop. Note that the test sets were not involved in model training or parameter tuning.

**Predictive performance measures**

The performances of the machine learning methods were evaluated in the test sets using three metrics, including log loss, area under the receiver operating characteristic curve (ROC-AUC) and area under the precision recall curve (PR-AUC). Log loss compares the predicted probabilities against the true labels. The receiver operating characteristic curve, which plots the sensitivity (*i.e.* recall) against (1- specificity), is a very widely used approach to evaluate predictive performances in biomedical applications. The precision-recall curve on the hand plots the precision (*i.e.* positive predictive value) against the sensitivity (recall). Since precision depends on the overall proportion of positive labels, the PR-AUC is also dependent on such proportions. Davies et al.[37] suggested that the PR curve may give more informative comparisons when working with imbalanced data.

**Identifying important genes and pathways**

We also performed analyses to reveal the genes which contribute the most to the prediction model. For elastic net, we extracted the genes with non-zero coefficient in at least one cross-validation fold, and the resulting genes were subject to an over-representation analysis (ORA) (using hypergeometric tests) to reveal the pathways involved. For RF and GBM, feature importance was computed using built-in functions in sklearn based on Gini importance (i.e. the average decrease in node impurity). We then performed a gene-set enrichment analysis (GSEA[38]) based on the genes together with their respective importance scores (the highest score across three folds was taken). For SVM and DNN, there is a lack of widely adopted importance measures, so we focused on the rest of ML methods in this part. Pathway analyses were conducted by the web-based program WebGastalt[39]. Four pathway databases were considered in our analyses, including KEGG, PANTHER, Reactome and Wikipathways.

**External validation by testing for enrichment of psychiatric drugs considered for clinical trials**
We then performed additional analyses to assess if the drugs with high predicted probabilities from our machine learning models are indeed good candidates for repositioning. Briefly, we tested whether the drugs with *no* known indication for the disease but high predicted probabilities are more likely to be included in clinical trials.

In the first step, we filtered off drugs that are known to be indicated for the disease as derived from ATC and MEDI-HPS. This is because we are mainly interested in repositioning *other* drugs of unknown therapeutic potential, and that the labels of drug indication (from ATC or MEDI-HPS) have already been utilized in the ML prediction steps. Next, we extracted a list of drugs that were included in clinical trials for schizophrenia as well as depression and anxiety disorders. The list was derived from clincialTrial.gov and we downloaded a pre-complied version from https://doi.org/10.15363/thinklab.d212.

We then tested for enrichment of those drugs listed in clinicalTrial.gov among the top repositioning results. We performed an enrichment analysis of "drug-sets", similar to a gene-set analysis approach widely used in bioinformatics[40]. We performed one-tailed *t*-tests to assess if the predicted probabilities (derived from machine learning models) are significantly higher for psychiatric drugs considered in clinical trials.

**RESULTS**

**Predictive performance comparison**
*Unweighted analysis*
The average predictive performances (averaged over three folds) of different machine learning methods are listed in Table 1. When considering log loss as the criteria of interest, SVM gave the best result overall, though EN showed the best performance in one of the four datasets. DNN and EN showed quite similar predictive performances. RF and GBM were slightly worse than other methods, but the difference was small.

When ROC-AUC was considered as the performance metric, SVM and EN gave similar performances. SVM outperformed EN in the schizophrenia datasets, while EN showed better results in the other two datasets. The performance of DNN was worse than that of SVM and EN, although the differences were not large. The two tree-based methods performed worse especially in the depression/anxiety datasets.

We then considered PR-AUC, which is more sensitive to imbalanced data, as the measure of predictive performance. SVM was the best-performing method. EN and DNN followed with very similar performances. Consistent with other performance measures, GBM and RF did not perform as well in the depression/anxiety datasets, but the performance was more comparable for the schizophrenia datasets.

*Weighted analysis*
Compared with unweighted analysis, we observed improvements in predictive performance for several methods including GBM, RF and deep learning. SVM and EN performed similarly in general. Considering ROC-AUC, deep learning performed the best for depression and anxiety disorders, while RF and EN showed highest ROC-AUC for schizophrenia. SVM achieved the best PR-AUC and log-loss compared to other ML approaches.

**Enrichment for psychiatric drugs considered in clinical trials**
We further tested whether the top repositioning results are enriched for drugs included in clinical trials for psychiatric disorders. As shown in Table 2, we observed significant enrichment of such drugs for both schizophrenia and depression/anxiety disorders across *all* methods in the weighted analysis. In addition, most results in the unweighted analysis were also significant. This external validation provides further support to the usefulness of our approach in identifying new repositioning opportunities.

**Identifying contributing genes and pathways**
Supplementary Tables 1-4 show the top genes as identified by variable importance measures (for RF and GBM) and regression coefficients (for EN). The enriched pathways are shown in Table 3 and Supplementary tables 5-12. Since the number of genes involved is large, we only highlighted a few top enriched pathways here. Interestingly, steroid and cholesterol biosynthesis are among the most significantly enriched pathways for drugs against schizophrenia and depression/anxiety. Notably, abnormalities in the hypothalamic-pituitary-adrenal (HPA) axis have long been suggested as one of the key pathological mechanisms underlying depression[41]. The steroid (cortisol) synthesis inhibitor metyrapone has been shown to be effective for depression in a double-blind randomized controlled trial (RCT)[42] and other studies (reviewed in ref.[43]), although another trial failed to show any benefits[44]. Antidepressants have also been shown to regulate glucocorticoid receptor functioning *in vivo*. On the other hand, neuroactive steroids may be implicated in the pathophysiology of schizophrenia[45]. Cholesterol biosynthesis, including regulation by sterol regulatory element-binding protein (SREBP), was frequently top-listed in our pathway analysis. Antipsychotics and some antidepressants are associated with metabolic syndrome and weight gain, and previous *in vitro* and *in vivo* studies have shown lipogenic effects of these drugs as controlled by SREBP transcription factors[46,47]. Interestingly, some studies showed lower cholesterol may be associated with suicidality[48], depressive symptoms[49,50], and poorer cognition in schizophrenia[51], but these findings are controversial. Whether pathways related to cholesterol synthesis may play a role in the therapeutic effects of psychotropic drugs remain a topic for further investigation. Some other pathways are also worth mentioning. For example, IGF signaling pathway was significantly enriched under antidepressants. IGF-I has been reported to improve depression and anxiety symptoms in clinical samples[52], and showed antidepressant-like effects in animal models (e.g. ref.[53,54]). The 5-HT$_3$ signaling pathway was also top-listed under antipsychotics. 5-HT$_3$ has been proposed as a new drug target and improvements in negative and cognitive symptoms have been reported in clinical trials[55].

**Top repositioning hits and literature support from previous studies**
Table 4 show some of the selected top repositioning candidates with literature support, which will also be discussed below. Tables 5 and 6 show the top 15 repositioning hits by DNN and SVM respectively. Predictions with the weighted ML models were presented here due to better overall predictive performances. More detailed tables showing the top 100 hits for each ML method in both unweighted and weighted analyses are presented in Supplementary Tables 13-16. Note that drugs that are known to be indicated for these disorders according to ATC or MEDI-HPS were excluded from the lists. We noted overlap in top hits derived from different machine learning methods, but some repositioning candidates are unique to one or few ML approaches. This suggests that employing a diverse set of ML methods may be advantageous in "learning" different potential repositioning candidates. We will chiefly focus on candidates from the top 15 hits for each ML method in the exposition below.

*Repositioning candidates for depression/anxiety disorders*

Regarding depression and anxiety disorders, many of the top results are antipsychotics, such as trifluoperazine, perphenazine, fluphenazine and thioridazine, among others. Antipsychotics have long been used for the treatment for depression[56]. In earlier studies, phenothiazines (a class of antipsychotic to which many of our top hits belong) was observed to produce similar anti-depressive effects as tricyclic antidepressants[57]. Due to the risk of extra-pyramidal side-effects, typical antipsychotics are less commonly used these days and second-generation (atypical) antipsychotics are more often prescribed. Meta-analyses have shown that atypical antipsychotics are effective as adjunctive or primary treatment for depression[58,59]. Antipsychotics are also commonly prescribed for severe depressive episodes with psychotic symptoms.

A few other drugs on the lists are also worth mentioning. Cyproheptadine (top-listed by SVM, RF, GBM and EN) is a 5-$HT_2$ receptor antagonist and was shown to improve depression in a small cross-over trial[60]. It was also reported that the drug reduced the neuropsychiatric side-effects of the antiviral therapy efavirenz, including depressive and anxiety symptoms[61]. Chlorcyclizine belongs to the phenylpiperazine class and numerous antidepressants and antipsychotics also belong to this class[62]. Pizotifen, listed by EN, is a 5-$HT_{2A/2C}$ antagonist which was shown to possess antidepressant effects in a double-blind RCT[63]. DNN and EN have identified histone deacetylase (HDAC) inhibitors including trichostatin A and vorinostat as top repositioning hits for depression/anxiety and schizophrenia. HDAC have been implicated in the pathogenesis of psychiatric disorders including depression, as reviewed by Fuchikami et al.[64]. HDAC inhibitors have been reported to produce antidepressant-like effects in animal models[65,66], although no clinical trials on psychiatric disorders were available. Interestingly, in a recent study which employed gene-set analysis on *de novo* mutations to uncover repositioning opportunities, HDAC inhibitors were highlighted as candidates for schizophrenia and other neurodevelopmental disorders[67].

Another candidate was tetrandrine, a calcium channel blocker top-listed by DNN, RF, GBM and SVM. Tetrandrine demonstrated antidepressant-like effects in mice[68] in forced swimming and tail suspension tests. The drug also increased the concentration of 5-hydroytrytamine (5-HT) and norepinephrine in mice treated with reserpine or chromic mild stress, and raised the levels of brain-derived neurotrophic factor (BDNF) in the latter case[68].

*Repositioning candidates for schizophrenia*

With regards to repositioning results for schizophrenia, some of the hits are antidepressants, such as protriptyline, maprotiline and clomipramine, among others. Antidepressants are frequently prescribed in schizophrenia patients due to possibility of comorbid depression or obsessive-compulsive disorder[69]. In meta-analyses antidepressants were also found to improve negative symptoms of schizophrenia[70,71]. For the antidepressants on the list, maprotiline (listed by RF, GBM, EN, SVM) has been reported to improve negative symptoms in chronic schizophrenia patients[72] as an adjunctive treatment. Other drug clomipramine (listed by DNN, GBM, EN) has been shown to ameliorate not only obsessive-compulsive but also overall schizophrenic symptoms in patients with comorbid disorders[73]. Interestingly, the mood stabilizer valproate was also listed among the top (by DNN and SVM). Valproate may improve clinical response when added to antipsychotics, although the evidence is mainly based on open RCTs[74]. The EN algorithm also "re-discovered" spiperone, an antipsychotic not listed in ATC or MEDI-HPS, as one of the top repositioning hits.

Several other drugs less well-known for psychiatric disorders are also worth mentioning. The selective estrogen receptor modulator raloxifene (listed by DNN and EN) was shown to improve schizophrenia symptom scores in double-blind RCTs of post-menopausal women[75,76]. Another drug nordihydroguaiaretic acid (listed by DNN, GBM, SVM) has antioxidant properties[77] and may be useful in combating oxidative stress in schizophrenia[78]. Pioglitazone, top-ranked by DNN, belongs to the class of thiazolidinediones and has anti-diabetic and anti-inflammatory properties. Although this drug was withdrawn due to unexpected adverse effects on the liver, our finding suggested that other thiazolidinediones may be useful for schizophrenia. Indeed, another drug of the same class known as pioglitazone has been shown to improve negative symptoms in schizophrenia patients in a double-blind RCT[79]. Another RCT also showed improvements in depressive symptoms [80]. Tretinoin (listed by DNN) is a retinoid and retinoid dysfunction has been linked to schizophrenia[81,82]. Clinical trials with another retinoid (bexarotene) showed some benefits of the drug as an add-on agent in schizophrenia. Again retinoid signaling was implicated for schizophrenia in a recent study on drug repositioning leveraging *de novo* mutations[67]. Felodipine (listed by GBM) is a calcium channel blocker and GWAS on schizophrenia and bipolar disorder have revealed many genes related to calcium channels[83,84]; a recent study also suggested concomitant use of CCB and antipsychotics may be more beneficial than antipsychotics alone[85].

*Some hits from the unweighted analysis*

The top repositioning candidates from *un*weighted analysis for each ML method are listed in supplementary tables 13-16. There were a number of overlaps with the candidates from the weighted analysis. Here we highlight a few prioritized drugs (that have not been mentioned above) with literature support. Aspirin (acetylsalicylic acid) is a non-steroidal anti-inflammatory agent (listed by SVM), which has been shown to improve schizophrenia symptoms in a recent meta-analysis of RCTs[86]. Genistein is a phytoestrogen and can bind to estrogen receptors[87]. An animal study showed that genistein may possess anti-dopaminergic actions[88]; interestingly, clinical studies have shown potential therapeutic benefits of estrogens on schizophrenia[86].

The EN algorithm identified metformin as one of the top repositioning hits for depression/anxiety disorders. A study in Taiwan reported that the risk of depression in diabetic patients was reduced by ~60% for those given metformin with sulfonylurea[89]. Another study reported improved depressive symptoms and cognitive functions for patients with comorbid diabetes and depression[90]. Another drug apigenin, top-listed by GBM, was supported by a number of *in vitro* and animal studies for possible antidepressant-like and anxiolytic effects[91]. A clinical trial of oral chamomile (which was standardized to contain 1.2% of apigenin) showed benefits for anxiety and depression[92,93].

**Discussions**

In this study, we have presented a repositioning approach by predicting drug indications based on expression profiles. We employed and compared five state-of-the-art machine learning methods to perform predictions. We also observed that the top repositioning hits are enriched for psychiatric drugs considered for clinical trials and that many hits are backed up by evidence form animal or clinical studies, supporting the validity of our approach.

Concerning the performance of different machine learning classifiers, we have employed five methods in total, and all but one (EN) are non-linear classifiers. SVM is a kernel-based learning approach that is widely used in bioinformatics. On the other hand, deep learning methods (such as DNN) that are based on the principles of representation learning[94] have witnessed rapid advances in the last decade, especially in the field of computer vision. While potentially powerful, current successful applications typically require very large datasets for training, and we suspect that the relatively modest sample size ($N = 3478$) of our dataset may have limited DNN to achieve the optimal predictive ability. We have used at most two hidden layers in view of the moderate sample size, and the complexity of the network may be increased with larger samples, although larger samples would lead to greater computational costs. This study shows that deep learning can achieve reasonable performance in drug repurposing, and indeed DNN achieved the best ROC-AUC for depression/anxiety disorders in the weighted analysis. Given the rapid growth in the area, deep learning approaches might be worthy of further investigations. While logistic regression with EN is a linear classifier, it performed well overall though lagging behind SVM. The performances of the two tree-based methods (RF and GBM) were largely comparable with other methods in the weighted analysis, although they were less satisfactory without weighting. Notwithstanding the differences in predictive performances, different algorithms are based on diverse model assumptions and principles, and as shown above, methods with slightly lower predictive accuracy may still reveal useful repositioning candidates that are of different mechanisms of actions.

To the best of our knowledge, this is the first study to employ a comprehensive array of machine learning methods on drug expression profiles for drug repositioning of any particular disease; it is also the first application in psychopharmacology. This is also the first work to suggest an ML approach to explore the molecular mechanisms underlying drug actions. In a related work, Aliper et al. made use of the drug transcriptome to predict drug classes *e.g.* drugs for neurological diseases, drugs for cardiovascular diseases, anti-cancer agents etc.[20]. Here our focus is different and perhaps clinically more relevant in that we directly identify repositioning opportunities for a particular disorder. In addition, we concentrated on the study of psychiatric disorders, which was not explicitly considered in Aliper et al.[20] or other previous works. Interestingly, DNN was reported to be the best performing method in their study. However their study[20] and the present work are not directly comparable as the outcomes studied are different and the evaluation metrics also differ. F1 score was used in Aliper et al.[20] (although the choice of a cut-off probability for classification was not explicitly stated) while we used ROC-AUC, PR-AUC and log loss as performance indicators.

Here we aim to provide a proof-of-concept example showing that the application of machine learning methodologies on drug expression profiles may help to identify candidates for repositioning, particularly for psychiatric disorders. The approach is intuitive and also highly flexible. Nevertheless, given the variety of methods and rapid advances in machine learning and computational drug repurposing, there is still room for improvement. Firstly, we only consider drug indications and the drug-induced transcriptomic changes in our prediction model. This makes the method very flexible and widely applicable to any compounds or drugs for which an expression profile is available. The use of drug transcriptome evades the need of specifying targets and knowing the mechanisms of actions, and the approach may even be applicable to a mixture of chemical ingredients as may be the case for herbal medicines. However, it is possible that our methods may be further improved by incorporating other information such as drug targets and chemical structure, if such information is available. As for the prediction algorithm, in our dataset the number of positive labels is small. We tackled this problem by adjusting the weighting of positive and negative instances and indeed found improvements for several ML approaches. Other methodologies to account for imbalanced data are also possible[95], and this may be a topic for further explorations. We covered five commonly used algorithms here but this coverage is obviously not complete; further studies may benefit from the use of more advanced or recently developed learning methods. We also notice that there is an ongoing effort to expand the coverage of CMap[96], and that the study with updated full data and documentations have just been released. We are planning to further explore the current framework in the expanded dataset.

It is reassuring to observe that many repositioning hits are supported by previous studies and the results enriched for psychiatric drugs considered in clinical trials. However, we stress that further well-designed pre-clinical and clinical studies are necessary before the any results can be brought into clinical practice. We have also made use of ML methods to explore potentially important genes and pathways that may contribute to treatment effects. However, the results require further experimental validations. Computational approaches for repositioning and explorations of drug mechanisms, such as ML-based methods, provide a cost-effective and systematic way to assess and prioritize drug candidates, and might help to reduce the high failure rates in drug development. Given the rising cost in developing a new drug, even a minute reduction in failure rate will represent large savings in absolute terms. Further work might involve combining the current approach with other computational and experimental methods to further improve the accuracy of drug repositioning.

It is widely acknowledged that drug development in psychiatry has become stagnant for some years, and that traditional approaches to drug discovery has not been as successful as anticipated. On the other hand, the past few years have seen an extremely rapid development in ML methods and applications; in this regard, we hope that this study will open a new avenue for drug repositioning/discovery, and stimulate further research to bridge the gap between ML and biomedical applications especially drug development. The list of repositioning candidates might also serve as a useful resource for researchers and clinicians working on schizophrenia as well as depression and anxiety disorders, which are illnesses very much in need of new therapies.

**Supplementary Information is available at** https://drive.google.com/open?id=1TyYXm4Vtcc0jB76xHzGeLKonax2ixy4a

**Acknowledgements**
This work is partially supported by the Lo-Kwee Seong Biomedical Research Fund and a Direct Grant from the Chinese University of Hong Kong. We thank Mr. Carlos Kwan-long Chau for help in pathway analysis, and Mr. AO Fu-Kiu Kelvin and Mr. WONG Yui-Hang Harris for assistance in literature search. We also thank Professor Stephen K.W. Tsui and the Hong Kong Bioinformatics Centre for computing support.

**Financial Disclosures**   The authors declare no competing financial interests.

Table 1   Average predictive performance of different machine learning models across four datasets in *un*weighted (top) and weighted analysis (bottom)

**Unweighted analysis**

*Average Log Loss*

|  | MEDI-HPS Depression/Anxiety | ATC antidepressants | MEDI-HPS Schizophrenia | ATC antipsychotics |
|---|---|---|---|---|
| SVM | **0.1188** | 0.0943 | **0.1018** | **0.0895** |
| EN | 0.1249 | **0.0916** | 0.1097 | 0.0954 |
| DNN | 0.124 | 0.0948 | 0.1111 | 0.0992 |
| GBM | 0.1293 | 0.1018 | 0.1157 | 0.1039 |
| RF | 0.1294 | 0.1002 | 0.1155 | 0.1013 |

*Average ROC-AUC*

|  | MEDI-HPS Depression/Anxiety | ATC antidepressants | MEDI-HPS Schizophrenia | ATC antipsychotics |
|---|---|---|---|---|
| SVM | **0.7141** | 0.7619 | **0.7705** | **0.7755** |
| EN | 0.725 | **0.779** | 0.7515 | 0.7681 |
| DNN | 0.6952 | 0.7456 | 0.7533 | 0.7604 |
| GBM | 0.6536 | 0.6042 | 0.7172 | 0.7433 |
| RF | 0.6315 | 0.6390 | 0.7036 | 0.7501 |

*Average PR-AUC*

|  | MEDI-HPS Depression/Anxiety | ATC antidepressants | MEDI-HPS Schizophrenia | ATC antipsychotics |
|---|---|---|---|---|
| SVM | **0.2026** | **0.1485** | **0.2973** | **0.3379** |
| EN | 0.1372 | 0.1008 | 0.1586 | 0.2087 |
| DNN | 0.1447 | 0.0877 | 0.1577 | 0.2156 |
| GBM | 0.0910 | 0.0417 | 0.1426 | 0.1528 |
| RF | 0.1193 | 0.0639 | 0.1677 | 0.1703 |

**Weighted analysis**

*Average Log Loss*

|  | MEDI-HPS Depression/Anxiety | ATC antidepressants | MEDI-HPS Schizophrenia | ATC antipsychotics |
|---|---|---|---|---|
| SVM | **0.1189** | **0.0934** | **0.1022** | **0.0898** |
| EN | 0.5803 | 0.5344 | 0.5028 | 0.5112 |
| DNN | 0.1309 | 0.0990 | 0.1308 | 0.1098 |
| GBM | 0.1281 | 0.1032 | 0.1114 | 0.0981 |
| RF | 0.1234 | 0.0943 | 0.1060 | 0.0939 |

*Average ROC-AUC*

|  | MEDI-HPS Depression/Anxiety | ATC antidepressants | MEDI-HPS Schizophrenia | ATC antipsychotics |
|---|---|---|---|---|
| SVM | 0.7198 | 0.7718 | 0.7731 | 0.7765 |
| EN | 0.6610 | 0.7394 | 0.7494 | **0.7997** |
| DNN | **0.7424** | **0.7979** | 0.7410 | 0.7576 |
| GBM | 0.7155 | 0.7578 | 0.7584 | 0.7794 |
| RF | 0.6890 | 0.7355 | **0.7843** | 0.7801 |

|     | *Average PR-AUC* |        |        |        |
| --- | ---    | ---    | ---    | ---    |
| SVM | **0.2017** | **0.1510** | **0.2980** | **0.3361** |
| EN  | 0.0751 | 0.0896 | 0.1520 | 0.2030 |
| DNN | 0.1796 | 0.1107 | 0.2278 | 0.2641 |
| GBM | 0.1800 | 0.1168 | 0.2697 | 0.2780 |
| RF  | 0.1771 | 0.1165 | 0.2721 | 0.2707 |

The learning algorithm with the best performance in each dataset (for each predictive performance measure) is marked in bold.

ROC-AUC: area under the curve (AUC) of the receiver operating characteristic (ROC) curve; PR-AUC: Average area under the curve (AUC) of the precision-recall (PR) curve.

SVM: support vector machines; EN: logistic regression with elastic net regularization; DNN: deep neural networks; RF: random forest; GBM, gradient boosted machines with trees.

MEDI-HPS: MEDication Indication - High Precision Subset; ATC: Anatomical Therapeutic Chemical classification.

The drug indications (treated as the outcome variable in prediction) are defined for two kinds of disorders (schizophrenia as well as depression and anxiety disorders) using two different sources (MEDI-HPS and ATC), hence a total of four datasets. Please refer to the main text for details.

Table 2    Enrichment for psychiatric drugs included in clinical trials among the repositioning hits

| | Dataset | unweighted analysis | | weighted analysis | |
|---|---|---|---|---|---|
| | | *P*-value | *q*-value | *P*-value | *q*-value |
| SVM | MEDI-HPS Depression/Anxiety | **0.0014** | **0.0168** | **0.0011** | **0.0031** |
| | ATC antidepressants | **0.0180** | 0.0615 | **0.0039** | **0.0065** |
| | MEDI-HPS Schizophrenia | **0.0205** | 0.0615 | **0.0264** | **0.0299** |
| | ATC antipsychotics | **0.0098** | 0.0588 | **0.0084** | **0.0116** |
| DNN | MEDI-HPS Depression/Anxiety | **0.0105** | **0.0180** | **0.0009** | **0.0030** |
| | ATC antidepressants | 0.1369 | 0.1369 | **0.0017** | **0.0043** |
| | MEDI-HPS Schizophrenia | 0.0908 | 0.0991 | **0.0190** | **0.0238** |
| | ATC antipsychotics | **0.0237** | **0.0316** | **0.0021** | **0.0046** |
| EN | MEDI-HPS Depression/Anxiety | **0.0022** | **0.0128** | **0.0023** | **0.0046** |
| | ATC antidepressants | **0.0032** | **0.0128** | **0.0087** | **0.0116** |
| | MEDI-HPS Schizophrenia | **0.0294** | **0.0353** | **0.0315** | **0.0332** |
| | ATC antipsychotics | **0.0104** | **0.0180** | **0.0033** | **0.0060** |
| GBM | MEDI-HPS Depression/Anxiety | **0.0494** | 0.0988 | **0.0003** | **0.0025** |
| | ATC antidepressants | **0.0433** | 0.0988 | **0.0002** | **0.0025** |
| | MEDI-HPS Schizophrenia | 0.2283 | 0.2747 | **0.0269** | **0.0299** |
| | ATC antipsychotics | 0.2482 | 0.2747 | **0.0005** | **0.0025** |
| RF | MEDI-HPS Depression/Anxiety | 0.0651 | 0.1116 | **0.0005** | **0.0025** |
| | ATC antidepressants | 0.2518 | 0.2747 | **0.0007** | **0.0028** |
| | MEDI-HPS Schizophrenia | 0.1299 | 0.1949 | **0.0427** | **0.0427** |
| | ATC antipsychotics | 0.5232 | 0.5232 | **0.0063** | **0.0097** |

P-values < 0.05 and q-values <0.05 are in bold.

Table 3  Selected enriched pathways based on variable importance of genes in ML models

| Method | Name | #Gene | FDR |
|---|---|---|---|
| **ATC antidepressants and MEDI-HPS depression/anxiety (weighted analysis)** | | | |
| eNet-ORA_Reactome | Cholesterol biosynthesis | 7 | 1.78E-06 |
| eNet-ORA_Wikipathway | Sterol Regulatory Element-Binding Proteins (SREBP) signalling | 8 | 1.73E-04 |
| eNet-ORA_KEGG | Steroid biosynthesis - Homo sapiens (human) | 5 | 2.38E-04 |
| gbm_KEGG | Fat digestion and absorption - Homo sapiens (human) | 34 | 8.95E-03 |
| gbm_Panther | Insulin/IGF pathway-protein kinase B signaling cascade | 34 | 1.05E-02 |
| rf_Wikipathway | Mismatch repair | 9 | 1.38E-01 |
| rf_Wikipathway | ID signaling pathway | 16 | 1.58E-01 |
| rf_Wikipathway | Statin Pathway | 25 | 1.61E-01 |
| rf_Wikipathway | Photodynamic therapy-induced HIF-1 survival signaling | 37 | 1.63E-01 |
| eNet-ORA_Panther | TGF-beta signaling pathway | 4 | 1.70E-01 |
| gbm_Reactome | STING mediated induction of host immune responses | 11 | 2.33E-01 |
| rf_KEGG | Amino sugar and nucleotide sugar metabolism - Homo sapiens (human) | 40 | 2.43E-01 |
| rf_Wikipathway | Phase I biotransformations, non P450 | 7 | 2.51E-01 |
| rf_Wikipathway | Target Of Rapamycin (TOR) Signaling | 30 | 2.56E-01 |
| rf_KEGG | Alanine, aspartate and glutamate metabolism - Homo sapiens (human) | 28 | 2.67E-01 |
| **ATC antipsychotics and MEDI-HPS schizophrenia (weighted analysis)** | | | |
| rf_Wikipathway | Sterol Regulatory Element-Binding Proteins (SREBP) signalling | 64 | 0.00E+00 |
| eNet-ORA_Reactome | Cholesterol biosynthesis | 7 | 9.33E-05 |
| eNet-ORA_KEGG | Steroid biosynthesis - Homo sapiens (human) | 4 | 2.01E-03 |
| rf_Wikipathway | Statin Pathway | 25 | 2.51E-02 |
| eNet-ORA_Wikipathway | Fatty Acid Beta Oxidation | 5 | 2.98E-02 |
| eNet-ORA_Reactome | Asparagine N-linked glycosylation | 15 | 4.36E-02 |
| eNet-ORA_KEGG | Metabolic pathways - Homo sapiens (human) | 37 | 4.87E-02 |
| eNet-ORA_Reactome | Synthesis of UDP-N-acetyl-glucosamine | 3 | 8.11E-02 |
| gbm_Panther | 5HT3 type receptor mediated signaling pathway | 14 | 8.58E-02 |
| eNet-ORA_KEGG | Citrate cycle (TCA cycle) - Homo sapiens (human) | 3 | 8.96E-02 |
| gbm_Reactome | G1/S-Specific Transcription | 18 | 1.32E-01 |
| eNet-ORA_Reactome | Antigen Presentation: Folding, assembly and peptide loading of class I MHC | 4 | 1.32E-01 |
| gbm_Panther | Androgen/estrogene/progesterone biosynthesis | 9 | 1.40E-01 |
| rf_Wikipathway | Photodynamic therapy-induced unfolded protein response | 23 | 1.42E-01 |
| eNet-ORA_Reactome | COPII (Coat Protein 2) Mediated Vesicle Transport | 6 | 1.44E-01 |

We aggregated pathway analysis results from 4 databases, namely KEGG, Reactome, Panther and Wikipathways. Pathways that were highly similar were filtered. FDR, false discovery rate; eNet, elastic net; gbm, gradient boosted machine; rf, random forest; ORA, over-representation analysis, applicable to elastic net. Only results from weighted analysis are included here.

Table 4    Some literature-supported candidates selected from top hits derived from machine learning methods (known antipsychotics and antidepressants are not included in this list)

| Drug | Method | Relationship with disease |
|---|---|---|
| **Depression/anxiety** | | |
| Cyproheptadine | SVM, RF, GBM, EN | 5-HT2 receptor antagonist, improve depression in a small cross-over trial |
| Chlorcyclizine | DNN, RF, GBM, SVM | phenylpiperazine group to which many other antidepressants and antipsychotics belong |
| Pizotifen | EN | 5-HT2A/2C antagonist, positive result in an RCT |
| TrichostatinA/Vorinostat | DNN, EN | HDAC inhibitors may have antidepressant effects as shown in animal models |
| Tetrandrine | DNN, RF, GBM | CCB; antidepressant-like effects in mice; may increase 5-HT, NE and BDNF concentrations |
| Apigenin | GBM | Antidepressant and anxiolytic effects in animal models and in an RCT |
| Metformin | EN | may reduce depression risk among DM subjects |
| **Schizophrenia** | | |
| Valproate | DNN, SVM | open RCTs reported symptom improvement when used as adjunctive treatment |
| Raloxifene | DNN, EN | improve SCZ symptoms in an RCT of post-menopausal women |
| Nordihydroguaiaretic acid | DNN, GBM, SVM | antioxidant; oxidative stress implicated in SCZ |
| Pioglitazone | DNN | Another drug in the same class (pioglitazone) improved SCZ symptoms in RCT |
| Tretinoin | DNN | Retinoid; dysfunction in retinoid signaling may be implicated in SCZ |
| Felodipine | GBM | CCB; CCB added to antipsychotics may be beneficial |
| Aspirin | SVM | NSAID; may improve SCZ symptoms as shown in RC |
| Genistein | GBM | Phytoestrogen; animal model show possible anti-dopaminergic effects |

Please refer to the main text for detailed discussions and references. As a number of top results were known antipsychotics or antidepressants (please refer to the main text for details), these were not presented in the above table.

SCZ, schizophrenia; RCT, randomized controlled trial; HDAC, Histone deacetylases; CCB, calcium channel blocker; 5-HT, serotonin; NE, norepinephrine; BDNF, brain-derived neurotrophic factor; NSAID, non-steroidal anti-inflammatory drugs.

Table 5  Top 15 repositioning candidates derived from the deep learning approach

| ATC antidepressants | | | ATC antipsychotics | | |
|---|---|---|---|---|---|
| | Drug_cell-line | Pred_prob | | Drug_cell-line | Pred_prob |
| 1 | homochlorcyclizine_MCF7 | 0.938 | 1 | metixene_MCF7 | 0.932 |
| 2 | thioridazine_PC3 | 0.866 | 2 | metixene_PC3 | 0.926 |
| 3 | pimethixene_MCF7 | 0.830 | 3 | mefloquine_PC3 | 0.916 |
| 4 | levomepromazine_PC3 | 0.829 | 4 | protriptyline_MCF7 | 0.907 |
| 5 | deptropine_PC3 | 0.761 | 5 | norcyclobenzaprine_PC3 | 0.906 |
| 6 | fluphenazine_PC3 | 0.756 | 6 | methylbenzethonium.chloride_MCF7 | 0.895 |
| 7 | trifluoperazine_PC3 | 0.756 | 7 | astemizole_MCF7 | 0.884 |
| 8 | cyproheptadine_PC3 | 0.736 | 8 | troglitazone_MCF7 | 0.867 |
| 9 | metixene_PC3 | 0.706 | 9 | clomipramine_MCF7 | 0.865 |
| 10 | tetrandrine_MCF7 | 0.702 | 10 | nortriptyline_MCF7 | 0.846 |
| 11 | norcyclobenzaprine_MCF7 | 0.700 | 11 | suloctidil_MCF7 | 0.835 |
| 12 | fluphenazine_HL60 | 0.695 | 12 | trimipramine_PC3 | 0.826 |
| 13 | flupentixol_PC3 | 0.691 | 13 | loperamide_MCF7 | 0.823 |
| 14 | norcyclobenzaprine_PC3 | 0.667 | 14 | perhexiline_MCF7 | 0.791 |
| 15 | loperamide_MCF7 | 0.667 | 15 | tetrandrine_MCF7 | 0.788 |

| MEDI-HPS depression/anxiety | | | MEDI-HPS schizophrenia | | |
|---|---|---|---|---|---|
| | Drug_cell-line | Pred_prob | | Drug_cell-line | Pred_prob |
| 1 | trichostatin A_HL60 | 0.999 | 1 | metixene_PC3 | 0.945 |
| 2 | fluphenazine_MCF7 | 0.864 | 2 | valproic.acid_MCF7 | 0.941 |
| 3 | promazine_PC3 | 0.840 | 3 | nordihydroguaiaretic.acid_MCF7 | 0.939 |
| 4 | metixene_PC3 | 0.817 | 4 | ciclosporin_MCF7 | 0.926 |
| 5 | orphenadrine_MCF7 | 0.791 | 5 | suloctidil_MCF7 | 0.922 |
| 6 | piperacetazine_PC3 | 0.757 | 6 | tetrandrine_MCF7 | 0.908 |
| 7 | homochlorcyclizine_MCF7 | 0.741 | 7 | desipramine_PC3 | 0.892 |
| 8 | fluphenazine_HL60 | 0.700 | 8 | loperamide_MCF7 | 0.887 |
| 9 | thioridazine_PC3 | 0.694 | 9 | tretinoin_MCF7 | 0.874 |
| 10 | tetrandrine_MCF7 | 0.682 | 10 | homochlorcyclizine_MCF7 | 0.864 |
| 11 | fluphenazine_PC3 | 0.654 | 11 | trimipramine_MCF7 | 0.857 |
| 12 | norcyclobenzaprine_PC3 | 0.609 | 12 | loperamide_PC3 | 0.838 |
| 13 | clomifene_PC3 | 0.600 | 13 | profenamine_PC3 | 0.829 |
| 14 | levomepromazine_PC3 | 0.595 | 14 | raloxifene_MCF7 | 0.799 |
| 15 | haloperidol_MCF7 | 0.580 | 15 | thiethylperazine_PC3 | 0.795 |

Table 6　Top 15 repositioning candidates derived from support vector machine

| | ATC antidepressants | | | ATC antipsychotics | |
|---|---|---|---|---|---|
| | Drug_cell-line | Pred_prob | | Drug_cell-line | Pred_prob |
| 1 | homochlorcyclizine_MCF7 | 0.764 | 1 | norcyclobenzaprine_MCF7 | 0.753 |
| 2 | cyproheptadine_MCF7 | 0.556 | 2 | protriptyline_MCF7 | 0.667 |
| 3 | thioridazine_MCF7 | 0.420 | 3 | methylbenzethonium.chloride_PC3 | 0.611 |
| 4 | trifluoperazine_MCF7 | 0.418 | 4 | astemizole_PC3 | 0.539 |
| 5 | chlorcyclizine_PC3 | 0.383 | 5 | metixene_MCF7 | 0.524 |
| 6 | levomepromazine_PC3 | 0.339 | 6 | metixene_PC3 | 0.517 |
| 7 | prochlorperazine_MCF7 | 0.323 | 7 | nordihydroguaiaretic.acid_HL60 | 0.488 |
| 8 | spiperone_PC3 | 0.311 | 8 | loperamide_MCF7 | 0.474 |
| 9 | perphenazine_MCF7 | 0.310 | 9 | suloctidil_MCF7 | 0.451 |
| 10 | pimethixene_MCF7 | 0.300 | 10 | norcyclobenzaprine_PC3 | 0.430 |
| 11 | trifluoperazine_PC3 | 0.300 | 11 | maprotiline_PC3 | 0.416 |
| 12 | CP-690334-01_PC3 | 0.264 | 12 | amitriptyline_PC3 | 0.391 |
| 13 | tetrandrine_MCF7 | 0.250 | 13 | quinisocaine_PC3 | 0.366 |
| 14 | loperamide_MCF7 | 0.232 | 14 | mefloquine_PC3 | 0.364 |
| 15 | chlorcyclizine_MCF7 | 0.228 | 15 | nortriptyline_MCF7 | 0.341 |

| | MEDI-HPS depression/anxiety | | | MEDI-HPS schizophrenia | |
|---|---|---|---|---|---|
| | Drug_cell-line | Pred_prob | | Drug_cell-line | Pred_prob |
| 1 | fluphenazine_MCF7 | 0.931 | 1 | nordihydroguaiaretic.acid_HL60 | 0.646 |
| 2 | thioridazine_PC3 | 0.780 | 2 | nordihydroguaiaretic.acid_MCF7 | 0.615 |
| 3 | fluphenazine_HL60 | 0.780 | 3 | methylbenzethonium.chloride_PC3 | 0.574 |
| 4 | homochlorcyclizine_MCF7 | 0.669 | 4 | quinisocaine_PC3 | 0.550 |
| 5 | fluphenazine_PC3 | 0.455 | 5 | nordihydroguaiaretic.acid_PC3 | 0.500 |
| 6 | cyproheptadine_MCF7 | 0.446 | 6 | norcyclobenzaprine_MCF7 | 0.420 |
| 7 | perphenazine_MCF7 | 0.422 | 7 | niclosamide_PC3 | 0.419 |
| 8 | nordihydroguaiaretic acid_HL60 | 0.356 | 8 | metixene_PC3 | 0.404 |
| 9 | thioridazine_HL60 | 0.345 | 9 | valproic.acid_PC3 | 0.404 |
| 10 | haloperidol_MCF7 | 0.331 | 10 | norcyclobenzaprine_PC3 | 0.401 |
| 11 | tetrandrine_MCF7 | 0.310 | 11 | suloctidil_MCF7 | 0.395 |
| 12 | nordihydroguaiaretic acid_PC3 | 0.310 | 12 | nortriptyline_MCF7 | 0.375 |
| 13 | metixene_PC3 | 0.308 | 13 | genistein_PC3 | 0.372 |
| 14 | cyproheptadine_PC3 | 0.291 | 14 | amitriptyline_MCF7 | 0.362 |
| 15 | metitepine_MCF7 | 0.287 | 15 | protriptyline_MCF7 | 0.362 |